\documentclass[9pt,twocolumn,twoside]{opticajnl}
\journal{opticajournal} 

\setboolean{shortarticle}{false}

\usepackage{lineno}
\usepackage[justification=justified]{caption}

\title{Advancing large-scale thin-film PPLN nonlinear photonics with segmented tunable micro-heaters}

\author[1,$\dagger$]{Xiaoting Li}
\author[1,$\dagger$]{Haochuan Li}
\author[1]{Zhenzheng Wang}
\author[1]{Zhaoxi Chen}
\author[2]{Fei Ma}
\author[1]{Ke Zhang}
\author[3,4,*]{Wenzhao Sun}
\author[1,5,*]{Cheng Wang}

\affil[1]{Department of Electrical Engineering, City University of Hong Kong, Kowloon, Hong Kong, China}
\affil[2]{School of Physics, Sun Yat-sen University, Guanzhou 510275, China}
\affil[3]{City University of Hong Kong (Dongguan), Dongguan, China.}
\affil[4]{Center of Information and Communication Technology, City University of Hong Kong Shenzhen Research Institute, Shenzhen, China.}
\affil[5]{State Key Laboratory of Terahertz and Millimeter Waves, City University of Hong Kong, Kowloon, Hong Kong, China.}

\affil[$\dagger$]{These authors contributed equally to this paper.}
\affil[*]{Corresponding author: wenzhao.sun@cityu-dg.edu.cn; cwang257@cityu.edu.hk}

\begin{abstract}
Thin-film periodically poled lithium niobate (TF-PPLN) devices have recently gained prominence for efficient wavelength conversion processes in both classical and quantum applications. However, the patterning and poling of TF-PPLN devices today are mostly performed at chip scales, presenting a significant bottleneck for future large-scale nonlinear photonic systems that require the integration of multiple nonlinear components with consistent performance and low cost. Here, we take a pivotal step towards this goal by developing a wafer-scale TF-PPLN nonlinear photonic platform, leveraging ultraviolet stepper lithography and an automated poling process. To address the inhomogeneous broadening of the quasi-phase matching (QPM) spectrum induced by film thickness variations across the wafer, we propose and demonstrate segmented thermal optic tuning modules that can precisely adjust and align the QPM peak wavelengths in each section. Using the segmented micro-heaters, we show the successful realignment of inhomogeneously broadened multi-peak QPM spectra with up to 57$\%$ enhancement of conversion efficiency. We achieve a high normalized conversion efficiency of 3802$\%$W$^{-1}$cm$^{-2}$ in a 6 mm long PPLN waveguide, recovering 84$\%$ of the theoretically predicted efficiency in this device. The advanced fabrication techniques and segmented tuning architectures presented herein pave the way for wafer-scale integration of complex functional nonlinear photonic circuits with applications in quantum information processing, precision sensing and metrology, and low-noise-figure optical signal amplification.
\end{abstract}

\setboolean{displaycopyright}{false} 

\begin{document}

\maketitle
	
\section{INTRODUCTION}
Thin-film periodically poled lithium niobate (TF-PPLN) devices, renowned for their strong optical nonlinearity and excellent light confinement, are essential nonlinear photonic building blocks for the next generation of optical communication and quantum information processing systems \cite{Boes_2023_Lithium}. Due to the substantially enhanced optical intensity in tightly confined waveguides, TF-PPLN wavelength convertors exhibit more than one order of magnitude higher normalized conversion efficiencies compared to their bulk counterparts \cite{Wang_2018_Ultrahighefficiency,Zhao_2020_Shallowetched,Rao_2019_Activelymonitored}. These highly efficient TF-PPLN waveguides have enabled many high-performance nonlinear devices, including resonator-based ultra-efficient wavelength converters \cite{Lu_2019_Periodically,Chen_2019_Ultraefficient}, broadband optical parametric amplifiers \cite{Ledezma_2022_Intense,Jankowski_2022_Quasistatic} and entangled photon-pair sources \cite{Zhao_2020_High,Xue_2021_Effect}. Moreover, TF-PPLN devices enjoy excellent compatibility with other on-chip functional photonic devices available on the thin-film lithium niobate (TFLN) platform, such as integrated EO modulators \cite{Wang_2018_Integrated,He_2019_Highperformance}, acousto-optic modulators \cite{Wan_2022_Highly}, frequency combs \cite{Wang_2019_Monolithic,He_2019_Selfstarting,Bruch_2021_Pockels}, as well as heterogeneously integrated lasers \cite{OpDeBeeck_2021_III} and photodetectors \cite{Desiatov_2019_Silicon,Guo_2022_Highperformance,Zhu_2023_Waveguide}. By now, this integration compatibility has empowered chip-scale nonlinear and quantum photonic systems with unprecedented performances, including efficient quantum squeezers \cite{Nehra_2022_Fewcycle,Stokowski_2023_Integrated}, femtosecond all-optical switches \cite{Guo_2022_Femtojoule}, octave-spanning optical parametric oscillators \cite{Ledezma_2023_Octavespanning}, and integrated Pockels lasers co-lasing at infrared and visible wavelengths \cite{Li_2022_Integrated}. Additionally, to facilitate the active control of quasi-phase-matching (QPM) wavelength, thermally tunable TF-PPLN waveguides with high tuning efficiencies have also been developed \cite{Liu_2022_Thermally}.\setlength{\parskip}{1pt}

Despite the remarkable achievements of TF-PPLN devices, their fabrication today still relies on waveguide patterning and crystal poling processes performed on individually centimeter-sized chips based on electron-beam lithography and manual poling. In recent years, wafer-scale fabrication techniques have been developed for TFLN devices with passive or electro-optic functionalities \cite{Luke_2020_Waferscale}; however, these processes have not yet been extended to include nonlinear photonic devices like TF-PPLN waveguides. This limitation persists mainly due to repeatability and throughput issues of the manual periodic poling processes. It is also technically challenging for an R$\&$D lab to reliably achieve high-quality nanoscale poling electrodes and accurate multi-layer alignment on a wafer scale. 

Another key challenge that arises when moving towards larger-scale integration and fabrication is the distortion of QPM spectra at extended PPLN waveguide lengths, since TF-PPLN waveguides are highly sensitive to variations in the optical waveguide dimensions due to their strong geometric dispersion. Among various factors, e.g. etching depth, top width and film thickness \cite{Fejer_1992_Quasiphasematched,Tian_2021_Effect,Xue_2021_Ultrabright,Santandrea_2019_Characterisation}, our previous study has concluded that film thickness variation is the predominant cause for the QPM spectrum degradation in TF-PPLN, which often leads to broadened or multi-peak QPM profiles and decreased conversion efficiencies \cite{Zhao_2023_Unveiling}. Our simulation results indicate that for 600 nm thick MgO-doped TF-PPLN waveguides, the QPM peak wavelength for second-harmonic generation (SHG) shifts by 6 nm when the film thickness changes by merely 1 nm. This is particularly problematic for a wafer-scale process where the film thickness variation across a lithium niobate on insulator (LNOI) wafer is typically ±10 nm, leading to significant distortion of the QPM spectrum within each PPLN device and inconsistent peak QPM wavelengths across different PPLN devices in a larger nonlinear photonic circuit.\setlength{\parskip}{1pt}

To address the QPM inhomogeneous broadening issue, it has been proposed that by fine-tuning the geometric parameters, an optimal noncritical phase-matching configuration can be achieved, rendering the PPLN waveguide less susceptible to variations in thickness \cite{Kuo_2022_Noncritical}. This method however requires a thicker film of 900 nm and a large etching depth, which is challenging in fabrication and not compatible with other commonly used devices in the TFLN platform. More recently, a novel approach has been introduced that leverages pre-fabrication mapping of the film thickness to design customized poling electrodes with domain inversion periods that are adapted to the local film thicknesses \cite{Chen_2023_Adapted}. This method effectively suppresses the QPM inhomogeneous broadening and enables a record-high overall conversion efficiency of 10,000$\%$W$^{-1}$ for PPLN waveguides. However, this technique relies on time-consuming two-dimensional thickness mapping and requires a unique poling electrode design for each chip, thus still face challenges in achieving high-throughput and cost-effective fabrication of future TF-PPLN nonlinear photonic circuits.

In this work, we tackle these challenges by developing a wafer-scale TF-PPLN nonlinear photonic platform with segmented thermal-optic (TO) tuning modules. We demonstrate reliable fabrication of TF-PPLN devices on a 4-inch TFLN wafer utilizing ultraviolet stepper lithography and an automated poling process. To counteract the inhomogeneous broadening effects resulting from film thickness variations across the wafer, we design and fabricate segmented micro-heaters that are capable of locally fine-tuning and aligning the QPM spectral peaks within each individual sections to achieve substantially enhanced wavelength conversion efficiencies. We show the successful recovery of a sinc-like QPM spectrum, with up to 57$\%$ improved peak SHG efficiency compared with the as-fabricated devices.

\section{DEVICE DESIGN AND OPERATION PRINCIPLE}

\begin{figure}[htb]
    \centering\includegraphics[width=0.48\textwidth]{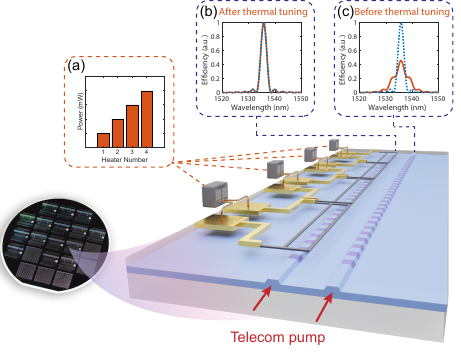}
    \caption{Schematic illustration of the wafer-scale PPLN optical waveguides featuring segmented micro-heaters. Insets: (a-b) Recovered QPM spectrum (b) after thermal tuning with segmented heating powers (a). (c) Broadened QPM spectrum due to thickness variation before thermal tuning.}
    \label{fig1}
\end{figure}
Figure \ref{fig1} presents a conceptual schematic of a wafer-scale nonlinear photonic platform based on TF-PPLN waveguides integrated with segmented TO tuning modules. Without micro-heaters, the QPM spectra of TF-PPLN waveguides typically see broadened or multi-peak profiles due to variations in film thickness and other geometric parameters (e.g., etching depth or top width), as shown in Fig. \ref{fig1}(c). By individually controlling the thermal power applied to each micro-heater [Fig. \ref{fig1}(a)], we can precisely adjust and align the QPM peaks to converge on the desired target peak wavelength, as shown in Fig. \ref{fig1}(b). The segmented micro-heaters essentially fine tune the effective film thickness in each section to enhance the global flatness of the TF-PPLN chip. Under this circumstance, the peak conversion efficiency of the PPLN waveguides could be recovered to approach the ideal level, depending on the remaining un-compensated film thickness variations. To some extent, this idea is similar to the historical differential heater designs in bulk LN nonlinear optics for compensation of material non-uniformity in non-congruently melting lithium niobate materials \cite{Nash_1970_Effect,Nash_1971_Influence}, and is also recently reported in Ti-diffused PPLN waveguides \cite{Zhao_2024_Tailored}.

\begin{figure}[htb]
    \centering\includegraphics{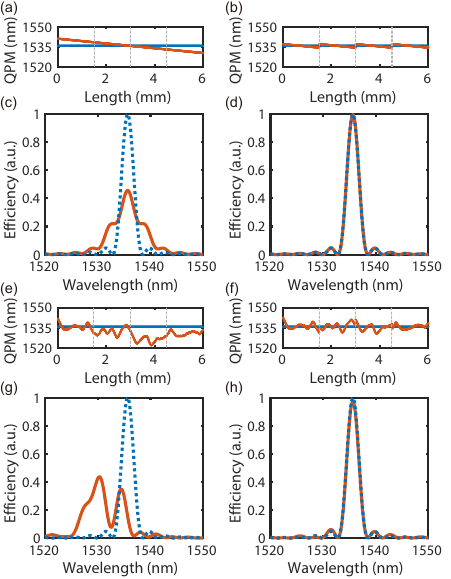}
    \caption{Simulated QPM spectra with (right column) and without (left column) segmented thermal tuning in the cases of linearly increasing film thickness (a-d) and a realistic thickness profile (e-h). (a) and (e) show the local QPM wavelengths along the TF-PPLN waveguides (red) compared with the target QPM wavelength (blue). (c) and (g) correspond to the simulated QPM spectra (red) in comparison with the ideal QPM spectrum (blue). (b) and (f) show the local QPM wavelength distributions after the center QPM wavelengths in each section are aligned by the segmented micro-heaters. (d) and (h) show the corresponding recovered QPM spectra by the micro-heaters.}
    \label{fig2}
\end{figure}
To validate our concept, we simulate the QPM spectra with thickness variation and those after optimized local thermal tuning, as illustrated in Fig. \ref{fig2}. Here we consider two scenarios: i) a hypothetical scenario where the film thickness linearly increases from the input to the output port; ii) a realistic scenario based on actually mapped thickness data from our recent research. In the first case, the film thickness linearly increases from 600 nm to 602 nm over a 6-mm device length, which corresponds to a linearly chirped peak QPM wavelength from 1529 nm to 1541 nm, as shown in Fig. \ref{fig2}(a). This leads to significant degradation in the peak conversion efficiency and deviation from the ideal QPM spectrum [Fig. \ref{fig2}(c), blue dashed curve denotes the ideal spectrum]. However, when we equip this inhomogeneously broadened TF-PPLN waveguide with four segmented TO tuning modules that align the center QPM wavelengths in each section [Fig. \ref{fig2}(b)], the normalized conversion efficiency is restored to 98$\%$ of the ideal value with a nearly sinc QPM spectrum, as shown in Fig. \ref{fig2}(d). To investigate the performance of our segmented tuning scheme in a more realistic scenario (second case), we use the mapped thickness data from our previous study [Fig. \ref{fig2}(e)] \cite{OpDeBeeck_2021_III}, which lead to a multi-peak QPM spectrum with a peak conversion efficiency $\sim$45.2$\%$ of the ideal case [Fig. \ref{fig2}(g)]. Similar to the first case, by aligning the local effective film thickness using micro-heaters, the normalized conversion efficiency could be enhanced by a factor of ~2.2, to 97$\%$ of the ideal case, as shown in Fig. \ref{fig2}(h). Moreover, the QPM spectrum is successfully recovered to a single main peak with a standard sinc profile. 

\section{DEVICE FABRICATION AND CHARACTERIZATION}

\begin{figure}[htb]
    \centering\includegraphics[width=0.45\textwidth]{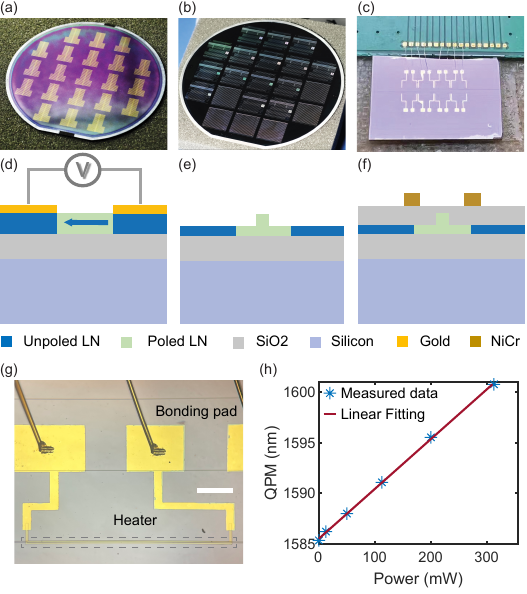}
    \caption{(a) The fabricated 4-inch LNOI wafer patterned with finger electrodes for periodic poling. (b) The wafer after periodic poling and patterning of optical waveguides. (c) The final cleaved TF-PPLN chip wire-bonded to a PCB. (d-f) Cross-sectional schematics of the fabrication process flow, specifically (d) high voltage poling; (e) optical waveguide formation using RIE etching; and (f) fabrication of the segmented micro-heaters. (g) Close-up microscope image of the fabricated segmented micro-heaters. Scale bar: 250 $\mu$m. (h) Measured QPM wavelength as a function of increasing heating power.}
    \label{fig3}
\end{figure}
We perform wafer-scale fabrication of the TF-PPLN devices using a process detailed in Fig. \ref{fig3}. The device is fabricated on a commercial 4-inch LNOI wafer supplied by NANOLN Ltd., comprising a 600 nm MgO-doped LN thin-film layer, a 2µm oxide buffer layer, and a 500µm silicon substrate. Firstly, the poling finger electrodes are patterned using an i-line UV stepper lithography (ASML), followed by thermal evaporation of nichrome (NiCr) and gold (Au) and a standard lift-off process, as shown in Fig. \ref{fig3}(a). Secondly, we employ a home-built automated probe station that is programmable to precisely position the probes on the poling electrodes sequentially and apply two 10-ms-long poling pulses, each reaching a peak voltage of 480V \cite{Zhao_2019_Optical,Nagy_2020_Submicrometer,Niu_2020_Optimizing}. Our homemade automated poling station is equipped with a two-dimensional precision control stage (Newport model 8742) for alignment of the sample in the X and Z directions, complemented by two three-dimensional translational stages that position the probes connected to the positive and negative poling electrodes. During the poling experiment, after ensuring proper contact between the first set of poling electrodes and the probes, the automated poling process is set to start, controlled by a LabVIEW program. During this process, the sample stage automatically moves down, advances forward, and moves up to initiate the next poling cycle, until the whole die containing 28 sets of electrodes are poled. This automation facilitates the reliable periodic poling of an entire 1.5 cm × 1.5 cm die without manual control or intervention, significantly reducing the workload of wafer-scale periodic poling. For additional operational details, please refer to the supplementary video illustrating the stage movement and poling sequence. Thirdly, after removal of all metal electrodes, a second aligned stepper lithography is carried out to define the patterns of optical waveguides in the poled regions. 800 nm thick AZ7908 photoresist is exposed at a dose of 230 mJ/cm$^{2}$ and developed for 60 s in FHD-5. The softbake and post-exposure bake temperatures are 90°C and 110°C for 60 s and 60 s, respectively. The exposed photoresist patterns are then transferred to the LN layer using an Ar$^{+}$-plasma-based reactive ion etching (RIE) process [Fig. \ref{fig3}(b)]. Subsequently, the fabricated TF-PPLN waveguides are cladded in silicon dioxide (SiO$_{2}$) using a plasma-enhanced chemical vapor deposition (PECVD) system. Fourthly, another two aligned photolithography processes are employed to fabricate the NiCr heaters in the vicinity of the optical waveguides, as well as the Au electrodes and bonding pads for wire bonding, similar to the process described in Ref. \cite{Liu_2022_Thermally}. Finally, the fabricated device undergoes cleaving and facet polishing to ensure good end-fire optical coupling. The Au electrode pads, consisting of 4 or 8 pairs of electrodes (each pair comprising one positive and one negative electrode), are wire-bonded to a printed circuit board (PCB) to facilitate independent control of each segmented micro-heater [Fig. \ref{fig3}(c)]. The gap between adjacent micro-heaters is 100 µm and the resistance of the 1.425-mm (for 6mm-long PPLN) and 1.1625-mm (for 1cm-long PPLN) long heater is $\sim$1000 $\Omega$ and $\sim$640 $\Omega$, respectively. The full device fabrication flow is illustrated in the cross-sectional schematics in Fig. \ref{fig3}(d)-(f). Figure \ref{fig3}(g) shows a close-up microscope image of the fabricated segmented micro-heaters. To evaluate the tuning efficiency of our segmented micro-heaters, uniformly increasing DC currents are applied simultaneously to all electrodes. As depicted in Fig. \ref{fig3}(h), the peak QPM wavelength exhibits a red shift in response to the incremental heating power, which indicates a thermal tuning efficiency of 50 pm/mW, which could be further improved by reducing the thermal power leakage using a suspended structure \cite{Liu_2022_Thermally}. During our testing over several days, we have not observed significant change or damage to the heaters or the wire bonds at heating powers up to 1.5 W.

\begin{figure}[htb]
    \centering\includegraphics{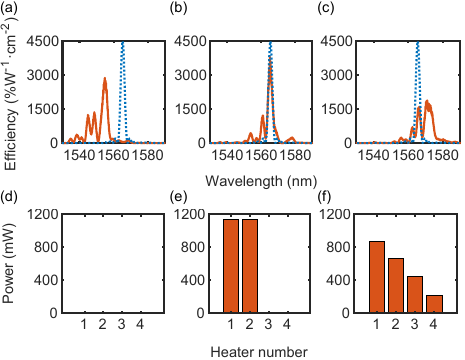}
    \caption{(a-c) Measured SHG intensities as functions of pump wavelengths for a 6mm TF-PPLN waveguide before applying tuning currents (a), after optimization of the heater powers (b), and with an arbitrary set of tuning parameters (c). (d-f) DC powers applied to each segmented micro-heater for the scenarios in (a-c) respectively.}
    \label{fig4}
\end{figure}

\begin{figure}[htb]
    \centering\includegraphics{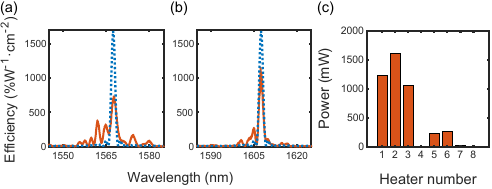}
    \caption{(a-b) Measured SHG intensities as functions of pump wavelengths for a 1cm TF-PPLN waveguide before (a) and after (b) applying tuning currents. (c) DC powers applied to each segmented micro-heater for the scenario in (b).}
    \label{fig4_2}
\end{figure}

Finally, we demonstrate the effective recovery of distorted QPM spectra using our segmented TO tuning modules, as shown in Figs. \ref{fig4}-\ref{fig4_2}. The TF-PPLN devices follow a similar design to our previous work \cite{Wang_2018_Ultrahighefficiency}, targeting SHG from telecom to near-visible bands. A telecom tunable light source (Santec TSL-550) is coupled into and out of the fabricated devices utilizing two optical lensed fibers. A fiber polarization controller is used to maintain a fundamental transverse-electric (TE) mode input. The measured SHG efficiency as a function of pump wavelength, also known as the QPM spectrum, is acquired by sweeping the input wavelength and simultaneously recording the output SHG power using a visible-band photodetector (Newport 1801). The on-chip SHG efficiency is obtained by carefully calibrating and de-embedding the visible and telecom coupling losses of the chip. For a 6mm long device in Fig. \ref{fig4}(a), the QPM profile without thermal tuning features three dominant peaks at 1545.1 nm, 1548.8 nm and 1554.9 nm. The highest peak conversion efficiency clearly falls short of the theoretical optimum ($\sim$64$\%$ compared with ideal QPM peak, blue dashed curve) due to the inhomogeneous broadening of the QPM spectrum. We subsequently apply DC currents to the four segmented micro-heaters integrated with this TF-PPLN waveguide. We first set the heating power of each micro-heater to 50$\%$ the maximum capacity to establish a baseline for tuning. By monitoring the QPM spectrum change when increasing/decreasing the power on each micro-heater, we obtain the qualitative tuning trend for each heater, which allows us to coarsely align the most prominent QPM peaks. We then fine-tune and optimize the QPM spectrum by iteratively adjusting the powers on each micro-heater [Fig. \ref{fig4}(e)] to achieve a single-main-peak QPM spectrum as depicted in Fig. \ref{fig4}(b). The measured peak second-harmonic (SH) conversion efficiency after thermal tuning is 3802$\%$W$^{-1}$cm$^{-2}$, which is increased by 32$\%$ from the initial value (2878$\%$W$^{-1}$cm$^{-2}$) and corresponds to ~84$\%$ the theoretical conversion efficiency (4500$\%$W$^{-1}$cm$^{-2}$). The remaining minor discrepancy from an ideal efficiency is mainly attributed to the small sub-peak at 1560.9 nm, which could not be merged into the main SHG peak in this particular set of device, possibly due to a larger thickness variation than expected at certain location of the chip. Moreover, it is also feasible to further re-shape the QPM spectra by applying another set of tuning currents, as exhibited in Fig. \ref{fig4}(c).   

We further fabricate and test a 1cm long TF-PPLN optical waveguide with 8 segmented micro-heaters, which ideally features a higher absolute conversion efficiency but is more prone to film thickness variations. As shown in Fig. \ref{fig4_2}(a), before the thermal tuning of micro-heaters, the QPM spectrum exhibits many unwanted sidelobes, which degrades the SHG conversion efficiency from the ideal value. Similar to the case above, by applying appropriate DC powers [as indicated in Fig. \ref{fig4_2}(c)], we achieved a 57$\%$ enhancement of peak SH conversion efficiency, with significantly suppressed sidelobes, as the measured QPM spectrum in Fig. \ref{fig4_2}(b) shows. The measured on-chip SHG efficiency for this device is 1153$\%$W$^{-1}$cm$^{-2}$ at optimized thermal tuning parameters (734$\%$W$^{-1}$cm$^{-2}$ before tuning). This value is $\sim$68$\%$ that of a device without inhomogeneous broadening (1700$\%$W$^{-1}$cm$^{-2}$), estimated by assuming the area underneath the QPM spectrum is invariant for inhomogeneous broadening. The remaining discrepancy from the simulated conversion efficiency is mainly due to insufficient poling depths in this 1-cm PPLN waveguides, which were fabricated from earlier, less optimized batch of TF-PPLN production. We also note that the areas beneath the QPM transfer functions before and after thermal tuning are consistent in both devices.

\section{CONCLUSIONS AND DISCUSSIONS}
In conclusion, we have demonstrated the wafer-scale production of TF-PPLN optical waveguides leveraging UV stepper lithography and an automated poling probe station. We address the degradation of conversion efficiency due to inhomogeneous film thickness by employing a segmented thermal tuning scheme. We show the successful recovery of single-peak QPM spectral profiles with up to 57$\%$ enhancement of the peak conversion efficiency, and achieve a highest normalized conversion efficiency of 3802$\%$W$^{-1}$cm$^{-2}$ in a 6 mm long device. Importantly, this is achieved without the need of pre-fabrication thickness mapping or design compensation, which is highly appealing for high-volume and low-cost wafer-scale production. The thermal tuning efficiency can be further enhanced by incorporating local air trenches to minimize heat leakage \cite{Liu_2022_Thermally}. LNOI wafers with better initial thickness variations are also expected to reduce the required heating powers in our devices. Even higher peak conversion efficiencies and better QPM spectral shapes could be achieved by implementing more thermal tuning modules and an automated control algorithm for optimizing the tuning parameters. This will enable faster searching for optimal working points, simultaneous control over multiple TF-PPLN devices, and real-time adaptation to environmental drifts. Our segmented heater design could also be combined with the adaptive poling method to compensate for the remaining inhomogeneous broadening effects and facilitate active tuning of QPM wavelengths. The scalable fabrication and tuning methodologies presented in this work mark an important step towards future large-scale nonlinear photonic integrated circuits with high efficiencies, versatile functionalities, and excellent reconfigurability, unlocking new opportunities for future quantum and classical photonic applications.

\begin{backmatter}
\bmsection{Funding} Research Grants Council, University Grants Committee (CityU 11204820, N$\_$CityU113/20); Croucher Foundation (9509005); Hong Kong PhD Fellowship Scheme (PF18-17958); National Natural Science Foundation of China (62105374).

\bmsection{Acknowledgments} The author would like to thank Dr. Wing-Han Wong and Mr. Hanke FENG for their kind help and guidance in the fabrication process.

\bmsection{Disclosures} The authors declare no conflicts of interest.

\bmsection{Data availability} Data underlying the results presented in this paper are not publicly available at this time but may be obtained from the authors upon reasonable request.

\end{backmatter}


\bibliography{micro-heater_try2}

\begin{thebibliography}{10}
\newcommand{\enquote}[1]{``#1''}

\bibitem{Boes_2023_Lithium}
A.~Boes, L.~Chang, C.~Langrock, \emph{et~al.}, \enquote{Lithium niobate photonics: {{Unlocking}} the electromagnetic spectrum,} {\protect\JournalTitle{Science}} \textbf{379}, eabj4396 (2023).

\bibitem{Wang_2018_Ultrahighefficiency}
C.~Wang, C.~Langrock, A.~Marandi, \emph{et~al.}, \enquote{Ultrahigh-efficiency wavelength conversion in nanophotonic periodically poled lithium niobate waveguides,} {\protect\JournalTitle{Optica}} \textbf{5}, 1438 (2018).

\bibitem{Zhao_2020_Shallowetched}
J.~Zhao, M.~R{\"u}sing, U.~A. Javid, \emph{et~al.}, \enquote{Shallow-etched thin-film lithium niobate waveguides for highly-efficient second-harmonic generation,} {\protect\JournalTitle{Optics Express}} \textbf{28}, 19669 (2020).

\bibitem{Rao_2019_Activelymonitored}
A.~Rao, K.~Abdelsalam, T.~Sjaardema, \emph{et~al.}, \enquote{Actively-monitored periodic-poling in thin-film lithium niobate photonic waveguides with ultrahigh nonlinear conversion efficiency of 4600 \%{{W}} {\textsuperscript{-1}} cm {\textsuperscript{-2}},} {\protect\JournalTitle{Optics Express}} \textbf{27}, 25920 (2019).

\bibitem{Lu_2019_Periodically}
J.~Lu, J.~B. Surya, X.~Liu, \emph{et~al.}, \enquote{Periodically poled thin-film lithium niobate microring resonators with a second-harmonic generation efficiency of 250,000\%/{{W}},} {\protect\JournalTitle{Optica}} \textbf{6}, 1455 (2019).

\bibitem{Chen_2019_Ultraefficient}
J.-Y. Chen, Z.-H. Ma, Y.~M. Sua, \emph{et~al.}, \enquote{Ultra-efficient frequency conversion in quasi-phase-matched lithium niobate microrings,} {\protect\JournalTitle{Optica}} \textbf{6}, 1244 (2019).

\bibitem{Ledezma_2022_Intense}
L.~Ledezma, R.~Sekine, Q.~Guo, \emph{et~al.}, \enquote{Intense optical parametric amplification in dispersion-engineered nanophotonic lithium niobate waveguides,} {\protect\JournalTitle{Optica}} \textbf{9}, 303 (2022).

\bibitem{Jankowski_2022_Quasistatic}
M.~Jankowski, N.~Jornod, C.~Langrock, \emph{et~al.}, \enquote{Quasi-static optical parametric amplification,} {\protect\JournalTitle{Optica}} \textbf{9}, 273 (2022).

\bibitem{Zhao_2020_High}
J.~Zhao, C.~Ma, M.~R{\"u}sing, and S.~Mookherjea, \enquote{High {{Quality Entangled Photon Pair Generation}} in {{Periodically Poled Thin-Film Lithium Niobate Waveguides}},} {\protect\JournalTitle{Physical Review Letters}} \textbf{124}, 163603 (2020).

\bibitem{Xue_2021_Effect}
G.-T. Xue, X.-H. Tian, C.~Zhang, \emph{et~al.}, \enquote{Effect of thickness variations of lithium niobate on insulator waveguide on the frequency spectrum of spontaneous parametric down-conversion*,} {\protect\JournalTitle{Chinese Physics B}} \textbf{30}, 110313 (2021).

\bibitem{Wang_2018_Integrated}
C.~Wang, M.~Zhang, X.~Chen, \emph{et~al.}, \enquote{Integrated lithium niobate electro-optic modulators operating at {{CMOS-compatible}} voltages,} {\protect\JournalTitle{Nature}} \textbf{562}, 101--104 (2018).

\bibitem{He_2019_Highperformance}
M.~He, M.~Xu, Y.~Ren, \emph{et~al.}, \enquote{High-performance hybrid silicon and lithium niobate {{Mach}}{\textendash}{{Zehnder}} modulators for 100 {{Gbit}} s-1 and beyond,} {\protect\JournalTitle{Nature Photonics}} \textbf{13}, 359--364 (2019).

\bibitem{Wan_2022_Highly}
L.~Wan, Z.~Yang, W.~Zhou, \emph{et~al.}, \enquote{Highly efficient acousto-optic modulation using nonsuspended thin-film lithium niobate-chalcogenide hybrid waveguides,} {\protect\JournalTitle{Light: Science \& Applications}} \textbf{11}, 145 (2022).

\bibitem{Wang_2019_Monolithic}
C.~Wang, M.~Zhang, M.~Yu, \emph{et~al.}, \enquote{Monolithic lithium niobate photonic circuits for {{Kerr}} frequency comb generation and modulation,} {\protect\JournalTitle{Nature Communications}} \textbf{10}, 978 (2019).

\bibitem{He_2019_Selfstarting}
Y.~He, Q.-F. Yang, J.~Ling, \emph{et~al.}, \enquote{Self-starting bi-chromatic {{LiNbO}} {\textsubscript{3}} soliton microcomb,} {\protect\JournalTitle{Optica}} \textbf{6}, 1138 (2019).

\bibitem{Bruch_2021_Pockels}
A.~W. Bruch, X.~Liu, Z.~Gong, \emph{et~al.}, \enquote{Pockels soliton microcomb,} {\protect\JournalTitle{Nature Photonics}} \textbf{15}, 21--27 (2021).

\bibitem{OpDeBeeck_2021_III}
C.~Op~De~Beeck, F.~M. Mayor, S.~Cuyvers, \emph{et~al.}, \enquote{{{III}}/{{V-on-lithium}} niobate amplifiers and lasers,} {\protect\JournalTitle{Optica}} \textbf{8}, 1288 (2021).

\bibitem{Desiatov_2019_Silicon}
B.~Desiatov and M.~Lon{\v c}ar, \enquote{Silicon photodetector for integrated lithium niobate photonics,} {\protect\JournalTitle{Applied Physics Letters}} \textbf{115}, 121108 (2019).

\bibitem{Guo_2022_Highperformance}
X.~Guo, L.~Shao, L.~He, \emph{et~al.}, \enquote{High-performance modified uni-traveling carrier photodiode integrated on a thin-film lithium niobate platform,} {\protect\JournalTitle{Photonics Research}} \textbf{10}, 1338 (2022).

\bibitem{Zhu_2023_Waveguide}
S.~Zhu, Y.~Zhang, Y.~Ren, \emph{et~al.}, \enquote{Waveguide-{{Integrated Two}}-{{Dimensional Material Photodetectors}} in {{Thin}}-{{Film Lithium Niobate}},} {\protect\JournalTitle{Advanced Photonics Research}} \textbf{4}, 2300045 (2023).

\bibitem{Nehra_2022_Fewcycle}
R.~Nehra, R.~Sekine, L.~Ledezma, \emph{et~al.}, \enquote{Few-cycle vacuum squeezing in nanophotonics,} {\protect\JournalTitle{Science}} \textbf{377}, 1333--1337 (2022).

\bibitem{Stokowski_2023_Integrated}
H.~S. Stokowski, T.~P. McKenna, T.~Park, \emph{et~al.}, \enquote{Integrated quantum optical phase sensor in thin film lithium niobate,} {\protect\JournalTitle{Nature Communications}} \textbf{14}, 3355 (2023).

\bibitem{Guo_2022_Femtojoule}
Q.~Guo, R.~Sekine, L.~Ledezma, \emph{et~al.}, \enquote{Femtojoule femtosecond all-optical switching in lithium niobate nanophotonics,} {\protect\JournalTitle{Nature Photonics}} \textbf{16}, 625--631 (2022).

\bibitem{Ledezma_2023_Octavespanning}
L.~Ledezma, A.~Roy, L.~Costa, \emph{et~al.}, \enquote{Octave-spanning tunable infrared parametric oscillators in nanophotonics,} {\protect\JournalTitle{Science Advances}} \textbf{9}, eadf9711 (2023).

\bibitem{Li_2022_Integrated}
M.~Li, L.~Chang, L.~Wu, \emph{et~al.}, \enquote{Integrated {{Pockels}} laser,} {\protect\JournalTitle{Nature Communications}} \textbf{13}, 5344 (2022).

\bibitem{Liu_2022_Thermally}
X.~Liu, C.~Zhang, Y.~Pan, \emph{et~al.}, \enquote{Thermally tunable and efficient second-harmonic generation on thin-film lithium niobate with integrated micro-heater,} {\protect\JournalTitle{Optics Letters}} \textbf{47}, 4921 (2022).

\bibitem{Luke_2020_Waferscale}
K.~Luke, P.~Kharel, C.~Reimer, \emph{et~al.}, \enquote{Wafer-scale low-loss lithium niobate photonic integrated circuits,} {\protect\JournalTitle{Optics Express}} \textbf{28}, 24452 (2020).

\bibitem{Fejer_1992_Quasiphasematched}
M.~Fejer, G.~Magel, D.~Jundt, and R.~Byer, \enquote{Quasi-phase-matched second harmonic generation: Tuning and tolerances,} {\protect\JournalTitle{IEEE Journal of Quantum Electronics}} \textbf{28}, 2631--2654 (Nov./1992).

\bibitem{Tian_2021_Effect}
X.-H. Tian, W.~Zhou, K.-Q. Ren, \emph{et~al.}, \enquote{Effect of dimension variation for second-harmonic generation in lithium niobate on insulator waveguide [{{Invited}}],} {\protect\JournalTitle{Chinese Optics Letters}} \textbf{19}, 060015 (2021).

\bibitem{Xue_2021_Ultrabright}
G.-T. Xue, Y.-F. Niu, X.~Liu, \emph{et~al.}, \enquote{Ultrabright {{Multiplexed Energy-Time-Entangled Photon Generation}} from {{Lithium Niobate}} on {{Insulator Chip}},} {\protect\JournalTitle{Physical Review Applied}} \textbf{15}, 064059 (2021).

\bibitem{Santandrea_2019_Characterisation}
M.~Santandrea, M.~Stefszky, G.~Roeland, and C.~Silberhorn, \enquote{Characterisation of fabrication inhomogeneities in {{Ti}}:{{LiNbO}} {\textsubscript{3}} waveguides,} {\protect\JournalTitle{New Journal of Physics}} \textbf{21}, 123005 (2019).

\bibitem{Zhao_2023_Unveiling}
J.~Zhao, X.~Li, T.-C. Hu, \emph{et~al.}, \enquote{Unveiling the origins of quasi-phase matching spectral imperfections in thin-film lithium niobate frequency doublers,} {\protect\JournalTitle{APL Photonics}} \textbf{8}, 126106 (2023).

\bibitem{Kuo_2022_Noncritical}
P.~S. Kuo, \enquote{Noncritical phasematching behavior in thin-film lithium niobate frequency converters,} {\protect\JournalTitle{Optics Letters}} \textbf{47}, 54 (2022).

\bibitem{Chen_2023_Adapted}
P.-K. Chen, I.~Briggs, C.~Cui, \emph{et~al.}, \enquote{Adapted poling to break the nonlinear efficiency limit in nanophotonic lithium niobate waveguides,} {\protect\JournalTitle{Nature Nanotechnology}}  (2023).

\bibitem{Nash_1970_Effect}
F.~R. Nash, G.~D. Boyd, M.~Sargent, and P.~M. Bridenbaugh, \enquote{Effect of optical inhomogeneities on phase matching in nonlinear crystals,} {\protect\JournalTitle{Journal of Applied Physics}} \textbf{41}, 2564--2576 (1970).

\bibitem{Nash_1971_Influence}
M.~Okada and S.~Ieiri, \enquote{Influences of self-induced thermal effects on phase matching in nonlinear optical crystals,} {\protect\JournalTitle{IEEE Journal of Quantum Electronics}} \textbf{7}, 560--563 (1971).

\bibitem{Zhao_2024_Tailored}
B.-H. Jonas, V.~B. Felix, H.~Harald, and S.~Christine, \enquote{Tailored second harmonic generation in ti-diffused ppln waveguides using micro-heaters,} {\protect\JournalTitle{Optics Express}} \textbf{32}, 6876 (2024).

\bibitem{Zhao_2019_Optical}
J.~Zhao, M.~R{\"u}sing, and S.~Mookherjea, \enquote{Optical diagnostic methods for monitoring the poling of thin-film lithium niobate waveguides,} {\protect\JournalTitle{Optics Express}} \textbf{27}, 12025 (2019).

\bibitem{Nagy_2020_Submicrometer}
J.~T. Nagy and R.~M. Reano, \enquote{Submicrometer periodic poling of lithium niobate thin films with bipolar preconditioning pulses,} {\protect\JournalTitle{Optical Materials Express}} \textbf{10}, 1911 (2020).

\bibitem{Niu_2020_Optimizing}
Y.~Niu, C.~Lin, X.~Liu, \emph{et~al.}, \enquote{Optimizing the efficiency of a periodically poled {{LNOI}} waveguide using {\emph{in situ}} monitoring of the ferroelectric domains,} {\protect\JournalTitle{Applied Physics Letters}} \textbf{116}, 101104 (2020).

\end{thebibliography}

\bibliographyfullrefs{micro-heater_try2}


\end{document}